\documentstyle[epsf,preprint]{ptptex}
\notypesetlogo
\preprintnumber{CU-TP/00-09}
\title{Small Violation of Universal Yukawa Coupling and Neutrino Large Mixing }
\author{T. {\sc Teshima}\footnote{E-mail address: teshima@isc.chubu.ac.jp} 
        and T. {\sc Asai}}
\inst{Department of Applied Physics,  Chubu University\\
 Kasugai 487-8501, Japan}
\recdate{\today}
\abst{We assume the universal Yukawa coupling (democratic mass matrix) 
with small violations for quarks , charged leptons and neutrinos masses. We 
could reproduce the mass hierarchy for quark masses and $V_{\rm CKM}$ matrix 
elements precisely. We adopt the see-saw mechanism for the explanation of 
smallness of neutrino masses and introduce the right-handed Majorana neutrinos 
and  Dirac neutrinos. We assume the universal Yukawa coupling 
with small violations for Majorana and Dirac neutrinos. We can get the 
hierarchy of charged lepton masses and effective neutrino masses and the 
large mixing of neutrinos expressed in $V_{\rm NMS}$.} 
\begin{document}
\maketitle
\section{Introduction}
Super-Kamiokande experiment has confirmed the $\nu_\mu\leftrightarrow
\nu_\tau$ oscillation to be very large $\sin^22\theta_{\rm atm}\sim1$ and 
the range of mass parameter $\Delta m_{\rm atm}^2$ to be $(2-5)\times10^{-3}
{\rm eV}^2$ by their atmospheric neutrino experiments.\cite{SUPERKAMIOKANDEI} 
Solar neutrino experiments analysis by Super-Kamiokande collaboration gives 
a conclusion that the large MSW solution is favored those which suggests 
$\nu_\mu\leftrightarrow\nu_e$ oscillation is large $\sin^22\theta_\odot\sim1$ 
and $\Delta m_\odot^2 \sim 10^{-5}-10^{-4}{\rm eV}^2$.
\cite{SUPERKAMIOKANDEII}
In the framework of three-flavor neutrinos, we can put $\Delta m^2_{\rm atm}$ 
to $\Delta m^2_{23}$ and $\Delta m^2_{\odot}$ to $\Delta m^2_{12}$, 
and then can consider the mass hierarchy $m_1\simeq m_2\ll m_3$. In this 
three-flavor neutrino framework,  $\sin^22\theta_{\rm atm}=\sin^22\theta_{23}
\sim1$ and $\sin^22\theta_{\odot}=\sin^22\theta_{12}\sim1$ and the remaining 
mixing is restricted to be $\sin^22\theta_{13}<0.10$ by the CHOOZ experiment.
\cite{CHOOZ} 
\par
The quark mixing is expressed by a $V_{\rm MKS}$ matrix and the neutrino 
mixing by a $V_{\rm MNS}$ matrix.\cite{MNS} The question why the neutrino 
sector mixings are so large although the quark sector mixings are small is 
the current most challenging one. The neutrino masses measured in neutrino 
oscillation bring the next question why the neutrino masses are so small. 
In order to explain these questions, many works \cite{MODELS,BRANCO} have been 
proposed. Almost works have studied adopting the so-called Froggatt Nielsen 
mechanism \cite{FNM} that is induced from the spontaneous breaking of some 
family symmetry. For the smallness of neutrino mass, almost works use the 
see-saw mechanism\cite{SEESAW} in which mass of the light neutrino is 
suppressed by a large scale of some unified theory. 
\par
For the quark sector, there is the universal Yukawa coupling approach (
democratic mass matrix approach), which explain the mass hierarchy and small 
mixing of $V_{\rm MKS}$ matrix.\cite{UNIVERSAL,TESHIMA} Especially, our 
approach \citen{TESHIMA} could reproduced the numerical results of quark mass 
hierarchy and $V_{\rm CKM}$ matrix elements precisely by using the universal 
Yukawa coupling with small violations. This approach stands on the following 
scenario:
(1) The main mass hierarchies between $(u, c)$ and $t$ in $(u, c, t)$ sector 
and between $(d, s)$ and $b$ in $(d, s, b)$ sector are induced by the 
universality of Yukawa coupling. This feature is characterized by the 
diagonalization of universal Yukawa coupling (democratic mass matrix) 
$M_0=m\left( \begin{array}{ccc}
                      1&1&1\\
                      1&1&1\\
                      1&1&1
               \end{array}\right)
$
for 
quarks to the ${\rm diag}[0,0,3m]$ using the unitary matrix 
$U_0=\left( \begin{array}{ccc}
                      1/\sqrt{2}&-1/\sqrt{2}&0\\
                      1/\sqrt{6}&1/\sqrt{6}&-2/\sqrt{6}\\
                      1/\sqrt{3}&1/\sqrt{3}&1/\sqrt{3}
               \end{array}\right)
$. 
Mass hierarchies between $u$ and $c$ in 
$(u, c)$ and $d$ and $s$ in $(d, s)$ are caused by small violations added 
to the universal Yukawa coupling $M_0$. 
(2) The smallness of the mixing parameters 
in $V_{\rm CKM}$ is produced by the difference between  the small violations 
for the $(u, c, t)$ sector and $(d, s, b)$ sector, because the $V_{\rm CKM}$ 
is the product of unitary transformation $T$ for the $(u, c, t)$ sector and 
$T^{\dagger}$ for the $(d, s, b)$ sector modified by the small violation from 
$T_0$. 
\par 
For the charged lepton and neutrino sector, we adopt the same scenario 
as (1) in order to explain the mass hierarchy of charged lepton and neutrino 
masses.  However, the neutrino masses are very small compared with the 
charged lepton masses by the order as $m_{\nu_\tau}/m_{\tau}\sim10^{-11}$. (3) 
In order to explain the smallness of neutrino masses, we adopt the see-saw 
mechanism introducing the right-handed Majorana neutrino with very 
large masses. We assume that the Majorana neutrino masses are produced 
by the universal Yukawa coupling with small violations as other fermions.
\par
From this scenario, we can get the hierarchical charged lepton masses and 
the transformation matrix $T$ modified by the small violation from $T_0$. For
neutrino masses, the effective neutrino mass $M_{\rm eff}$ produced through 
the see-saw mechanism are expressed as $M_{\rm eff}=M_DM^{-1}_MM_D^t$, where 
$M_D$ and $M_M$ are the Dirac neutrino mass matrix and the Majorana neutrino 
mass matrix, respectively. Though these neutrino mass matrices $M_D$ and $M_M$ 
are democratic mass matrices in our scenario, the effective neutrino 
mass matrix $M_{\rm eff}$ could be almost diagonal if the small violations 
in $M_M$ satisfy some condition. If the effective neutrino mass matrix 
$M_{\rm eff}$ is almost diagonal, translating matrix $T$ for neutrino is 
almost unit matrix. Then the neutrino mixing matrix $V_{\rm MNS}\sim T_0$ 
and large neutrino mixing is realized. Recently an analysis \cite{BRANCO} 
using the same scenario as our present one is presented, but the pattern of 
violation parameters added to the democratic mass matrix is different from 
ours.   
\par
The condition realizing the large neutrino mixing does not depend on the 
detail of the model. The stability of the condition is also guaranteed with 
respect to radiative corrections. Our scenario uses the similarity between 
$(d,~s,~b)$ quarks and $(e,~\mu,~\tau)$ lepton, and the see-saw mechanism 
introducing the Majorana neutrino. Thus our approach has to assume the 
unified $SU(5)$ symmetry at the least. We can get the rather precise numerical 
rule of the violation parameters for quark sector, but not get the precise 
numerical result for lepton sector. In order to discuss about the generation 
symmetry, we have to get the more precise numerical data on the lepton sector.

\section{Violation of universal Yukawa coupling}
Usually, in order to generate the mass hierarchy of quarks and leptons, 
Froggatt and Nielsen \cite{FNM} mechanism are used. This mechanism assumes 
that an abelian horizontal symmetry $U(1)_X$ and higher dimensional 
operators involving one or several electroweak singlet scalar fields which 
acquire a vacuum expectation values breaking the horizontal symmetry at some 
large scale. In procedure using this mechanism, the pattern of mass 
hierarchy of quarks and leptons and mixings for these fermions are sensitive 
to the horizontal symmetry adopted and charges of the horizontal symmetry 
assigning to the fields concerned. 
\par
In the universal Yukawa coupling procedure, main mass hierarchy is produced 
by the universality of the Yukawa coupling (democratic mass matrix) and 
another mass hierarchy is produced by the small violations adding to the 
democratic mass matrix. This violation is just considered as the $SU(3)$ 
violation in hadron spectroscopy and hadron decay processes. This $SU(3)$ 
violation has been considered to be produced by 
the quark mass difference (violation from the $SU(3)$ symmetry) and quark 
dynamics. Similarly, the violations adding to democratic mass matrix are 
considered to be produced by some violation from a horizontal symmetry and 
some dynamics of quarks and leptons. Because the origin of violation is not 
clear at present, we have to treat these small violations as free parameters. 

\subsection{Quark sector}
We use the following quark mass matrices with small violations of the Yukawa 
coupling strength containing the phases, 
\begin{eqnarray}
M^q&=&\Gamma^q\left(\begin{array}{ccc}
                1&(1-\delta^q_1)e^{i\phi^q_1}&(1-\delta^q_2)e^{i\phi^q_2}\\
                (1-\delta^q_1)e^{-i\phi^q_1}&1&(1-\delta^q_3)e^{i\phi^q_3}\\
                (1-\delta^q_2)e^{-i\phi^q_2}&(1-\delta^q_3)e^{-i\phi^q_3}&1
                \end{array}\right),\hspace*{0.2cm}
                (q=u, d)\\
&&\hspace*{0.7cm}\delta^{u,d}_i\ll1,\ \phi^{u, d}_i\ll1.\ \ (i=1,2,3)\nonumber
\end{eqnarray}
These mass matrices are diagonalized by the unitary transformations $U_L
(\delta^q_1,~ \delta^q_2,~ \delta^q_3,$
$~ \phi^q_1,~ \phi^q_2,~ \phi^q_3)$ 
and $U_R(\delta^q_1,~ \delta^q_2,~ \delta^q_3,~ \phi^q_1,~ \phi^q_2,~ 
\phi^q_3)$ as the formulae:
\begin{eqnarray}
&&U_L(\delta^q_1,~ \delta^q_2,~ \delta^q_3,~ \phi^q_1,~ \phi^q_2,~ \phi^q_3)
M^qU_R^{-1}(\delta^q_1,~ \delta^q_2,~ \delta^q_3,~ \phi^q_1,~ \phi^q_2,~ 
\phi^q_3)=M^q_D,\ \ \ (q=u,\ d)\nonumber\\
&&M^u_D={\rm diag}[m_u,\ m_c,\ m_t],\ \ \ M^d_D={\rm diag}[m_d,\ m_s,\ m_b].
\end{eqnarray}
In the limit of $\delta^q_i\to0$ and $\phi^q_i\to0$, these mass matrices are 
diagonalized to ${\rm diag}[0,~0,~3\Gamma^q]$ by the unitary transformation 
$U_0$;
\begin{eqnarray}
&&U_0M^q(\delta^q_i\to0,\ \phi^q_i\to0)U^{-1}_0={\rm diag}[0, 0, 3\Gamma^q],\ \ \ q=u,\ d\nonumber\\ 
&&U_0=\left( \begin{array}{ccc}
                      \frac1{\sqrt{2}}&\frac{-1}{\sqrt{2}}&0\\
                      \frac1{\sqrt{6}}&\frac1{\sqrt{6}}&\frac{-2}{\sqrt{6}}\\
                      \frac1{\sqrt{3}}&\frac1{\sqrt{3}}&\frac1{\sqrt{3}}
               \end{array}\right).
\end{eqnarray}
In the present procedure, $\delta^q_i$ and $\phi^q_i$ are very small and then 
$U_L(\delta^q_1,~ \delta^q_2,~ \delta^q_3,~ \phi^q_1,~ \phi^q_2,~ \phi^q_3)$ 
and $U_R(\delta^q_1,~ \delta^q_2,~ \delta^q_3,~ \phi^q_1,~ \phi^q_2,~ 
\phi^q_3)$ have the form near to $U_0$. 
The CKM matrix representing the quark mixing is defined as
\begin{equation}
V_{\rm CKM}=U_L(\delta^u_1,~ \delta^u_2,~ \delta^u_3,~ \phi^u_1,~ \phi^u_2,~ 
            \phi^u_3)U^{\dagger}_L(\delta^d_1,~ \delta^d_2,~ \delta^d_3,~ 
            \phi^d_1,~ \phi^d_2,~ \phi^d_3).
\end{equation} 
\par
We have carried out this procedure numerically and gotten the precise results 
in previous work.\cite{TESHIMA} Here, we will only show those results. We 
adopted the following numerical data:
\begin{eqnarray}
&&\frac{m_u}{m_c}=0.0038\pm0.0025,\ \ \ \frac{m_d}{m_s}=0.050\pm0.035,
          \nonumber\\
&&\frac{m_c}{m_t}=0.0042\pm0.0013,\ \ \ \frac{m_d}{m_s}=0.038\pm0.019,
         \nonumber\\
&&\Gamma^u=\frac{m_t}{3},\ \ \ \Gamma^d=\frac{m_b}{3},\nonumber\\
&&V_{\rm CKM}=\left(\begin{array}{ccc}
          0.9747\mbox{--}0.9759&0.218\mbox{--}0.224&0.002\mbox{--}0.005\\
          0.218\mbox{--}0.224&0.9738\mbox{--}0.9752&0.032\mbox{--}0.048\\
          0.004\mbox{--}0.015&0.030\mbox{--}0.048&0.9988\mbox{--}0.9995
          \end{array}\right).
\end{eqnarray}
From these numerical data, we could get the results for the violation 
parameters:
\begin{eqnarray}
&&\delta^u_1=0.00001\mbox{--}0.0004,\ \ \delta^u_+\equiv\frac{\delta^u_2+
           \delta^u_3}2=0.0064\mbox{--}0.0125,\ \ \ \delta^u_-\equiv
           \delta^u_2-\delta^u_3=\pm(0.0\mbox{--}0.0043),\nonumber\\
&&\delta^d_1=0.001\mbox{--}0.015,\ \ \delta^d_+\equiv\frac{\delta^d_2+
           \delta^d_3}2=0.040\mbox{--}0.129,\ \ \ \delta^d_-\equiv
           \delta^d_2-\delta^d_3=\pm(0.038\mbox{--}0.006),\nonumber\\
&&\hspace*{3cm}\phi^d_+\equiv\frac{\phi^d_2+\phi^d_3}2=-4^{\circ}\mbox{--}
           -3^{\circ},\ \ \ \phi^d_-\equiv\phi^d_2-\phi^d_3=\pm(
           -1^{\circ}\mbox{--}0^{\circ}).
\end{eqnarray}
These parameters seems to have a power rule parameterized by only 2 
parameters, $\lambda$ and $\phi$, as
\begin{eqnarray}
&&\delta^u_1=\lambda^8, \ \ \delta^u_-=\lambda^6,\ \ \delta^u_+=
           \lambda^4,\nonumber\\
&&\delta^d_1=\lambda^4, \ \ \delta^d_-=\lambda^3,\ \ \delta^d_+=
           \lambda^2,\ \ \lambda\approx0.32,\nonumber\\
&&\phi_+\equiv\phi\approx-4^\circ,
\end{eqnarray}
where we used the running masses for $m_t$ and $m_b$ masses at the scale 
$\mu=1{\rm GeV}$.\cite{RUNNING} This very simple parameterization seems to 
give some suggestion to the flavor symmetry.  
\subsection{lepton sector}
\par
Now, we analyze the charged lepton and neutrino mass hierarchy and neutrino 
mixing matrix $V_{\rm MNS}$. Similarly as $d$ quark sector, the charged lepton 
mass matrix is expressed as 
\begin{equation}
M^l=\Gamma^l\left(\begin{array}{ccc}
                1&1-\delta^l_1&1-\delta^l_2\\
                1-\delta^l_1&1&1-\delta^l_3\\
                1-\delta^l_2&1-\delta^l_3&1
                \end{array}\right),\ \ \ 
\delta^l_i\ll1\ \  (i=1,2,3)
\end{equation}
where the phase factor is neglected at present analysis. For neutrino masses, 
we use the see-saw mechanism and introduce the Dirac 
neutrino $M^{\nu}_D$ and Majorana neutrino $M^{\nu}_M$. These neutrino masses 
produce the effective neutrino masses expressed as
\begin{eqnarray}
&&M^{\nu}_{\rm eff}=M^{\nu}_D{M^{\nu}_M}^{-1}(M^{\nu}_D)^t,\\
&&M^{\nu}_D=\Gamma^{\nu}_D\left(\begin{array}{ccc}
                1&1-\delta^{\nu}_1&1-\delta^{\nu}_2\\
                1-\delta^{\nu}_1&1&1-\delta^{\nu}_3\\
                1-\delta^{\nu}_2&1-\delta^{\nu}_3&1
                \end{array}\right),\ \   
\delta^{\nu}_i\ll1\ \ (i=1,2,3)\nonumber\\
&&M^{\nu}_M=\Gamma^{\nu}_M\left(\begin{array}{ccc}
                1-\Delta^{\nu}_1&1-\Delta^{\nu}_2&1-\Delta^{\nu}_3\\
                1-\Delta^{\nu}_2&1-\Delta^{\nu}_4&1-\Delta^{\nu}_5\\
                1-\Delta^{\nu}_3&1-\Delta^{\nu}_5&1
                \end{array}\right).\ \   
\Delta^{\nu}_i\ll1\ \ (i=1,2,3,4,5)\nonumber
\end{eqnarray}
For $M^\nu_M$, we add the breaking term to (1,1) and (2,2) element in order 
to keep the generality. 
\par 
The charged lepton mass matrix is diagonalized by the unitary transformation 
$U_L(\delta^l_1,~\delta^l_2,~\delta^l_3)$ and $U_R(\delta^l_1,~\delta^l_2,
~\delta^l_3)$ similarly to quark sector as 
\begin{eqnarray}
&&U_L(\delta^l_1,~ \delta^l_2,~ \delta^l_3)M^lU_R^{-1}(\delta^l_1,~ \delta^l_2,
~ \delta^l_3)=M^l_D,\ \ \ \nonumber\\
&&M^l_D={\rm diag}[m_e,\ m_\mu,\ m_\tau].
\end{eqnarray}
As the quark sector, $U_L(\delta^l_1,~ \delta^l_2,~ \delta^l_3)$ and $U_R
(\delta^l_1,~ \delta^l_2,~ \delta^l_3)$ have the form near to $U_0$. 
In fact, for the mass ratios for charged leptons, $\frac{m_e}{m_\mu}=
0.004836$, $\frac{m_\mu}{m_\tau}=0.05946\pm0.00001$, then the parameters  
$\delta^\nu_i$ and transformation matrix $U_L(\delta^l_i)$ are taken as
\begin{eqnarray}
&&\delta^l_1=0.002,\ \ \delta^l_2=0.137,\ \ \delta^l_3=0.113,\nonumber\\
&&U_L(\delta^l_i)=\left(\begin{array}{ccc}
        0.6726&-0.7363&0.0727\\
        0.4547&0.3339&-0.8256\\
        0.5837&0.5884&0.5594
        \end{array}\right).
\end{eqnarray} 
\par     
On the other hand, the neutrino mass matrix $M^\nu_{\rm eff}$ is diagonalized 
by the transformation matrix $U_L(\delta^\nu_1,~ \delta^\nu_2,~ \delta^\nu_3,
~\Delta^\nu_1,~ \Delta^\nu_2,
~ \Delta^\nu_3,~\Delta^\nu_4,~ \Delta^\nu_5)$ as 
\begin{eqnarray}
&&U_L(\delta^\nu_1,~ \delta^\nu_2,~ \delta^\nu_3,~\Delta^\nu_1,~ \Delta^\nu_2,
~ \Delta^\nu_3,~\Delta^\nu_4,~ \Delta^\nu_5)M^\nu_{\rm eff}U_L^{-1}
(\delta^\nu_1,~ \delta^\nu_2,~ \delta^\nu_3,~\Delta^\nu_1,~ \Delta^\nu_2,~ 
\Delta^\nu_3,~\Delta^\nu_4,~ \Delta^\nu_5)\nonumber\\
&&\hspace*{2cm}=M^\nu_{\rm diag},\ \ \ \nonumber\\
&&\hspace*{1cm}M^\nu_{\rm diag}={\rm diag}[m_{\nu_e},\ m_{\nu_\mu},\ 
m_{\nu_\tau}].
\end{eqnarray}
The $V_{\rm MNS}$ matrix is defined as 
\begin{equation}
V_{\rm MNS}=U^{\dagger}_L(\delta^l_1,~ \delta^l_2,~ \delta^l_3)U_L
            (\delta^\nu_1,~ \delta^\nu_2,~ \delta^\nu_3,~\Delta^\nu_1,~ 
            \Delta^\nu_2,~ \Delta^\nu_3,~\Delta^\nu_4,~\Delta^\nu_5).
\end{equation} 
Because $U_L(\delta^l_1,~\delta^l_2,~\delta^l_3)$ is nearly equal to $U_0$, 
if $U_L(\delta^\nu_1,~ \delta^\nu_2,~ \delta^\nu_3,~\Delta^\nu_1,~ 
\Delta^\nu_2,~ \Delta^\nu_3,~\Delta^\nu_4,~\Delta^\nu_5)$ is nearly equal to 
unit matrix, $V_{\rm MNS}$ matrix is nearly equal to $U^{\dagger}_0$, that is, 
the neutrino mixing is almost bi-maximal; $\nu_\mu\mbox{-}\nu_e$ mixing is 
maximal ($\theta_{12}=45^\circ$) and $\nu_\mu\mbox{-}\nu_\tau$ mixing is 
almost maximal ($\theta_{23}=35.3^\circ$). 
\par
We will consider the possibility that $U_L(\delta^\nu_1,~ \delta^\nu_2,~ 
\delta^\nu_3,~\Delta^\nu_1,~ \Delta^\nu_2,~ \Delta^\nu_3,~\Delta^\nu_4,
~\Delta^\nu_5)$ becomes nearly equal to unit matrix. This possibility is 
achieved in the case that $M^\nu_{\rm eff}$ is almost diagonal.
We calculate the $M^\nu_{\rm eff}$ using the Eq.(9) and then get the result;
\begin{eqnarray}
&&M^\nu_{\rm eff}=\frac{\Gamma^2_D}{\Gamma_M}\frac{1}{{\rm Det}_2(\Delta^\nu)
+{\rm Det}_3(\Delta^\nu)}
\left(\begin{array}{ccc}
M^\nu_{11}&M^\nu_{12}&M^\nu_{13}\\
M^\nu_{21}&M^\nu_{22}&M^\nu_{23}\\
M^\nu_{31}&M^\nu_{32}&M^\nu_{33}
\end{array}\right),\\
&&{\rm Det}_2(\Delta^\nu)=(\Delta^\nu_2-\Delta^\nu_3-\Delta^\nu_5)^2
             -(\Delta^\nu_1-2\Delta^\nu_3)(\Delta^\nu_4-
             2\Delta^\nu_5),\nonumber\\
&&{\rm Det}_3(\Delta^\nu)=2{\Delta^\nu_2}{\Delta^\nu_3}{\Delta^\nu_5}
             -{\Delta^\nu_1}{\Delta^\nu_5}^2-{\Delta^\nu_3}^2
             {\Delta^\nu_4},
             \nonumber\\
&&M^\nu_{ij}=M^\nu_{ji}={\rm Det}_2(\Delta^\nu)+\Delta M^\nu_{ij}(\Delta^\nu,
             \delta^\nu),\nonumber             
\end{eqnarray}
where ${\rm Det}_2(\Delta^\nu)$ and ${\rm Det}_3(\Delta^\nu)$ are order 2 
and 3 part of $\Delta^\nu_i$ in determinant of $M^\nu_M$ respectively, 
and $\Delta M^\nu_{ij}(\Delta^\nu,\delta^\nu)$ represent the term more than 
order 3 of $\Delta^\nu_i$ and $\delta^\nu_i$. It is stressed that the 
$M^\nu_{ij}$ contain the ${\rm Det}_2(\Delta^\nu)$ term commonly for all 
$(i,~j)$ elements. Because the order 2 term of $\Delta^\nu_i$ is usually 
larger than the terms more than order 3 of $\Delta^\nu_i$ and $\delta^\nu_i$, 
the mass matrix $M^\nu_{\rm eff}$ becomes nearly democratic mass matrix, 
\begin{equation}
M^\nu_{\rm eff}\approx\frac{\Gamma^2_D}{\Gamma_M}
\left(\begin{array}{ccc}
1&1&1\\
1&1&1\\
1&1&1
\end{array}\right),\ \ \ {\rm for}\ \ {\rm Det}_2(\Delta^\nu)\ne0.
\end{equation}
Then in this case, we cannot get the large mixing for neutrinos. However, 
if the term ${\rm Det}_2(\Delta^\nu)$ is exact 0, the situation is drastically 
changed
\begin{eqnarray}
&&M^\nu_{\rm eff}=\frac{\Gamma^2_D}{\Gamma_M}\frac{1}{{\rm Det}_3(\Delta^\nu)}
\left(\begin{array}{ccc}
\Delta M^\nu_{11}(\Delta^\nu,\delta^\nu)&\Delta M^\nu_{12}
(\Delta^\nu,\delta^\nu)&\Delta M^\nu_{13}(\Delta^\nu,\delta^\nu)\\
\Delta M^\nu_{21}(\Delta^\nu,\delta^\nu)&\Delta M^\nu_{22}
(\Delta^\nu,\delta^\nu)&\Delta M^\nu_{23}(\Delta^\nu,\delta^\nu)\\
\Delta M^\nu_{31}(\Delta^\nu,\delta^\nu)&\Delta M^\nu_{32}
(\Delta^\nu,\delta^\nu)&\Delta M^\nu_{33}(\Delta^\nu,\delta^\nu)
\end{array}\right),\nonumber\\
&& \hspace*{2.5cm} {\rm for}\ \ {\rm Det}_2(\Delta^\nu)=0
\end{eqnarray}
then the mass matrix $M^\nu_{\rm eff}$ can be the nearly diagonal mass 
matrix by choosing the small violations $\delta^\nu_i$ appropriately. 
\par 
We will search such small violations $\Delta^\nu_i$ and $\delta^\nu_i$ that 
the neutrino mass matrix $M^\nu_{\rm eff}$ becomes the nearly diagonal matrix, 
equally the $U_L(\delta^\nu_1,~\delta^\nu_2,~ \delta^\nu_3,~
\Delta^\nu_1,~ \Delta^\nu_2,~ \Delta^\nu_3,~$ $\Delta^\nu_4,~ \Delta^\nu_5)$ 
in Eq.(11) becomes nearly unit matrix; 
\begin{eqnarray}
&&(i,j\ne i) {\rm \ elements\ of\ } U_L(\delta^\nu_1,~ \delta^\nu_2,
~ \delta^\nu_3,~\Delta^\nu_1,~ \Delta^\nu_2,~ \Delta^\nu_3,~\Delta^\nu_4,
~ \Delta^\nu_5)<0.1,\\
&&
\begin{array}{l}
{\rm Det}_2(\Delta^\nu)=(\Delta^\nu_2-\Delta^\nu_3-\Delta^\nu_5)^2
             -(\Delta^\nu_1-2\Delta^\nu_3)(\Delta^\nu_4-
             2\Delta^\nu_5)
             =0,\\
{\rm Det}_3(\Delta^\nu)=2{\Delta^\nu_2}{\Delta^\nu_3}{\Delta^\nu_5}
             -{\Delta^\nu_1}{\Delta^\nu_5}^2-{\Delta^\nu_3}^2
             {\Delta^\nu_4}
             \ne0.
\end{array}                           
\end{eqnarray}
We can get the parameters satisfying the conditions Eqs.~(17),~(18). 
For example 
\begin{eqnarray}
&&\begin{array}{lllll}
\Delta^\nu_1=0.009,&\Delta^\nu_2=0.007,&\Delta^\nu_3=0.004,&
\Delta^\nu_4=0.005,&\Delta^\nu_5=0.002,\\
\delta^\nu_1=0.01,&\delta^\nu_2=0.02,&\delta^\nu_3=0.22,&&
\end{array}\nonumber\\
&&U_L(\delta^\nu_i,\Delta^\nu_i)=\left(\begin{array}{ccc}
           0.9931&-0.0991&-0.0627\\
           0.1016&0.9938&0.0444\\
           0.0579&-0.0505&0.9970
    \end{array}\right).\nonumber
\end{eqnarray}
For various values for the parameters $\Delta^\nu_i$ satisfying the condition 
Eq.~(18), there are many solutions for $\delta^\nu_i$ satisfying the condition 
Eq.~(17). We show the allowed points satisfying the condition Eq.~(17) in 
parameter space of $\delta^\nu_1$, $\delta^\nu_2$, $\delta^\nu_3$ in Fig.1.
The area of the circle on the point in parameter space is 
proportional to the numbers of the combinations of $\Delta^\nu_i$ 
satisfying the condition Eqs.~(17) and (18). Fig.~1(a) shows the case 
$\delta^\nu_1=0$, and Fig.~1(b) the case $\delta^\nu_1=0.025$, Fig.~1(c) the 
case $\delta^\nu_1=0.05$ and Fig.~1(d) the case $\delta^\nu_1=0.075$.
  
\begin{figure}
    \epsfxsize=10cm
  \centerline{\epsfbox{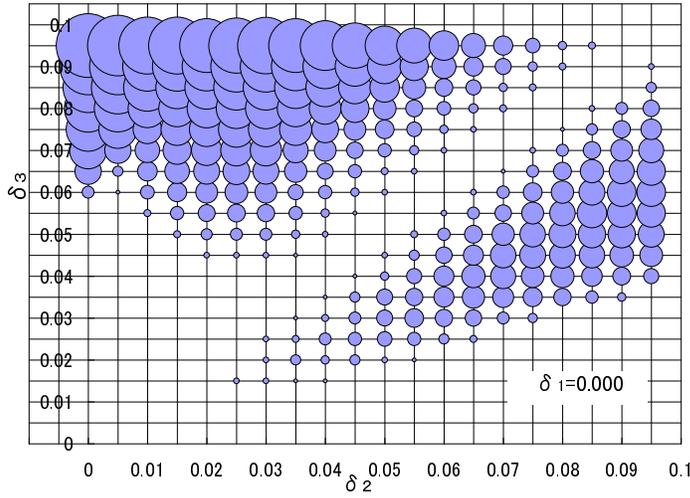}}
\caption{(a)\ Allowed points satisfying the condition Eq.~(17) in parameter 
plane ($\delta^\nu_2$, $\delta^\nu_3$) for $\delta^\nu_1=0.000$. For the size 
of circle, see the text}
\end{figure}   
\setcounter{figure}{0}
\begin{figure}
  \epsfxsize=10cm
  \centerline{\epsfbox{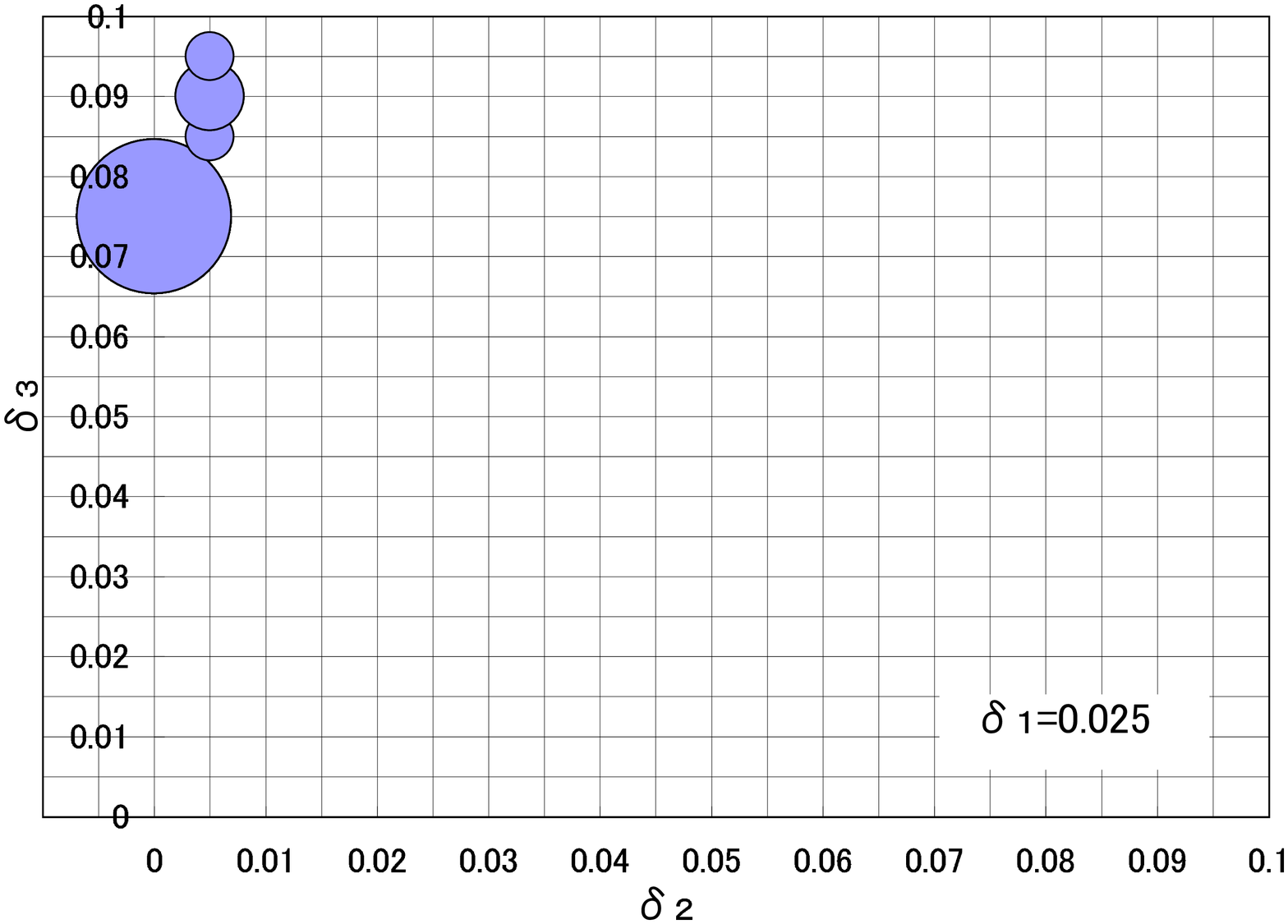}}
\caption{(b)\ Allowed points satisfying the condition Eq.~(17) in parameter 
plane ($\delta^\nu_2$, $\delta^\nu_3$) for $\delta^\nu_1=0.025$. For the size 
of circle, see the text}   
\end{figure}
\setcounter{figure}{0}   
\begin{figure}
   \epsfxsize=10cm
  \centerline{\epsfbox{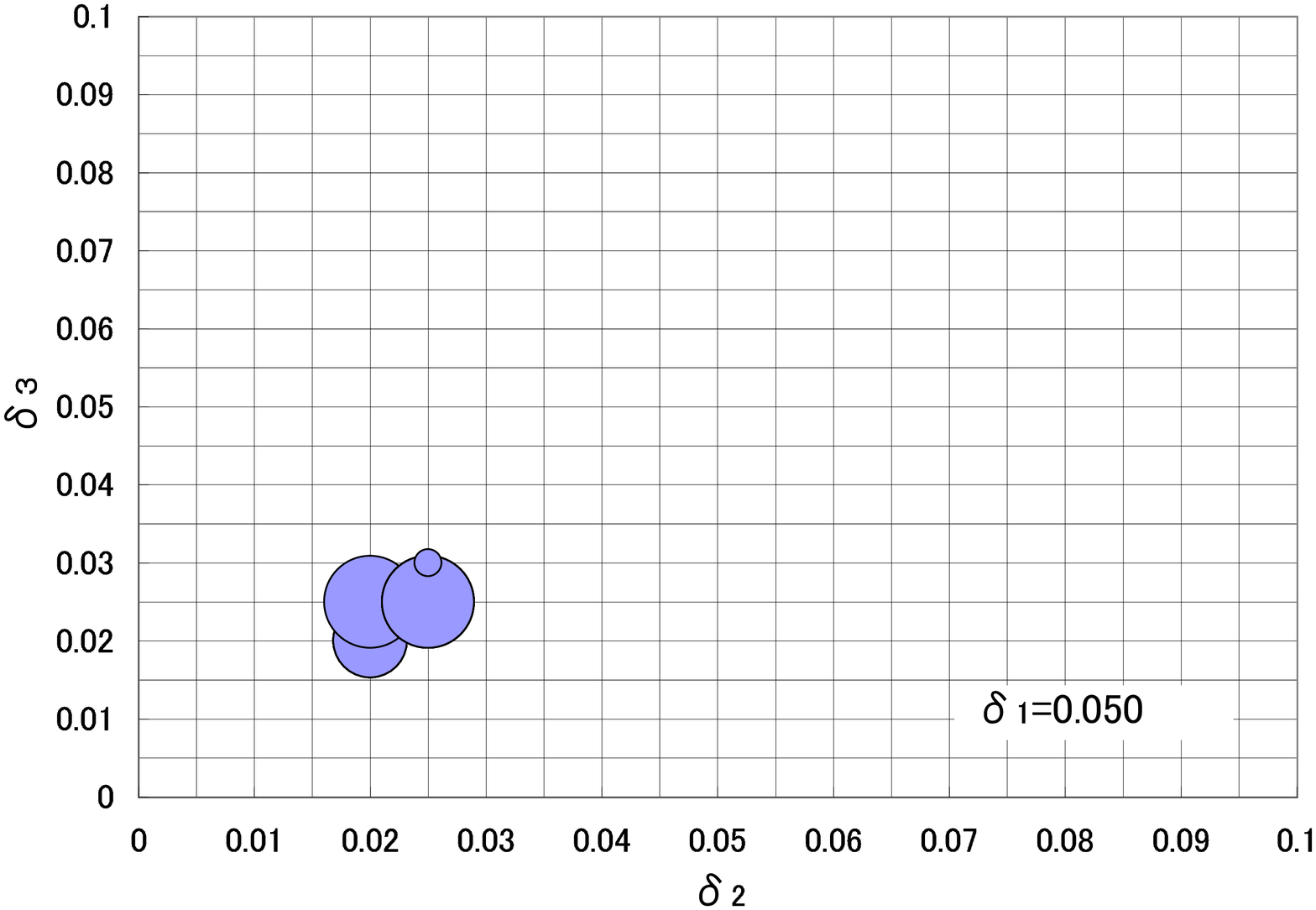}}
\caption{(c)\ Allowed points satisfying the condition Eq.~(17) in parameter 
plane ($\delta^\nu_2$, $\delta^\nu_3$) for $\delta^\nu_1=0.050$. For the size 
of circle, see the text}
\end{figure} 
\setcounter{figure}{0}   
\begin{figure}
  \epsfxsize=10cm
  \centerline{\epsfbox{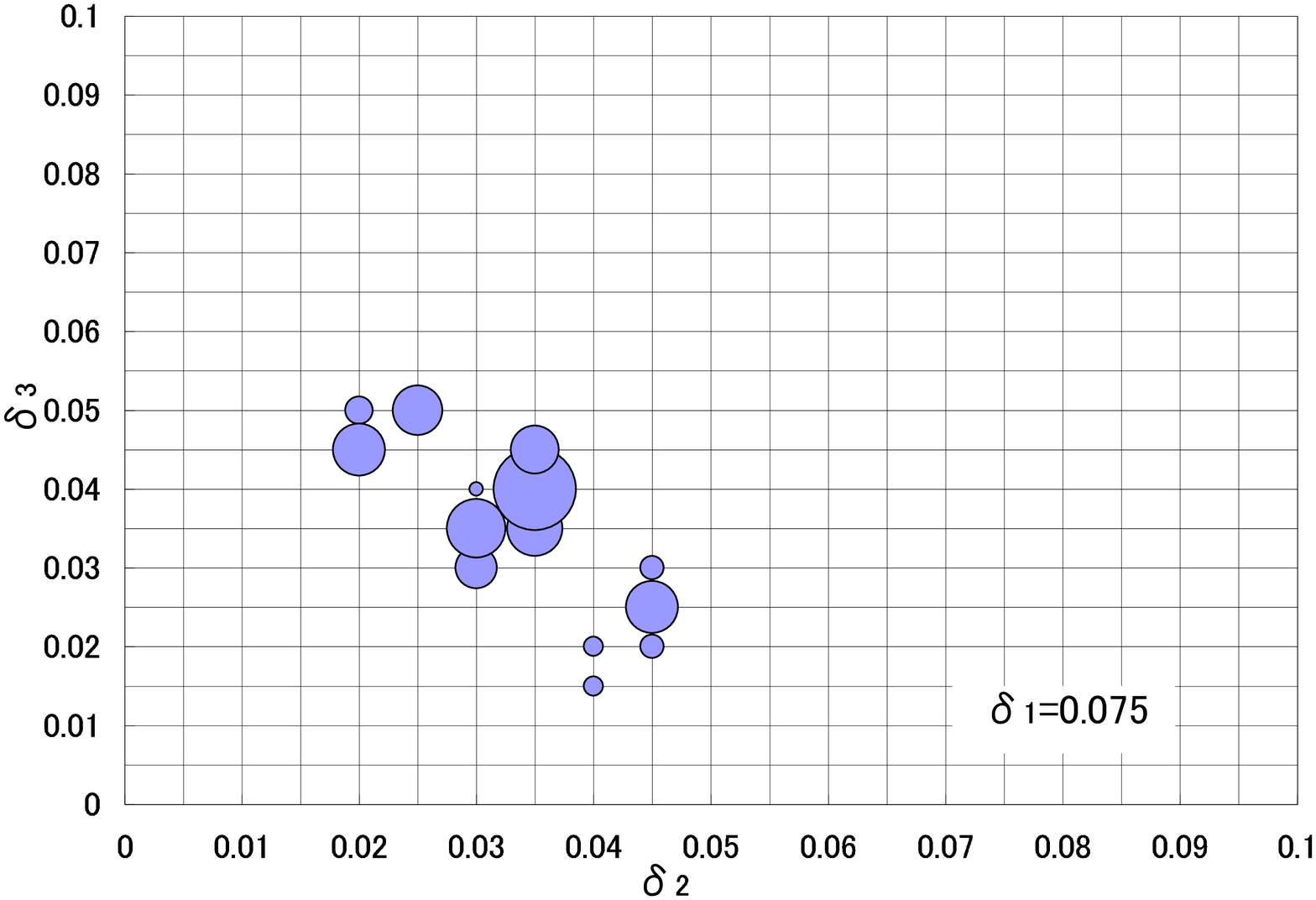}}
\caption{(d)\ Allowed points satisfying the condition Eq.~(17) in parameter 
plane ($\delta^\nu_2$, $\delta^\nu_3$) for $\delta^\nu_1=0.075$. For the size 
of circle, see the text}  
\label{fig1}
\end{figure}  

\par
The large mixing of neutrino is achieved by satisfying the condition Eq.~(18) 
for the violation parameters of $M^\nu_M$. Our present analysis is similar to 
the one of Ref.~\citen{BRANCO}, but the pattern of violation parameters added 
to the democratic mass matrix is different from theirs. Next work is to 
find a reason producing the condition Eq.~(18), but it seems difficult. 
We will study this problem in future articles. Here, we examine the stability 
of the condition. That is, although this condition is satisfied at a scale, 
there is no assurance that the condition is satisfied at other scale. However 
the renormalization effect relates $M^\nu_M(M_R)$ at the GUT scale $M_R$ to 
$M^\nu_M(M_Z)$ at the scale $M_Z$\cite{HABA} as;
\begin{eqnarray}
&&M^\nu_M(M_Z)=\left(\begin{array}{ccc}
             1-\epsilon_e&0&0\\
             0&1-\epsilon_\mu&0\\
             0&0&1
             \end{array}\right)^{-1}
             M^\nu_M(M_R)
             \left(\begin{array}{ccc}
             1-\epsilon_e&0&0\\
             0&1-\epsilon_\mu&0\\
             0&0&1
             \end{array}\right)^{-1}\nonumber\\
&&\epsilon_{e,\mu}=1-\sqrt{\frac{I_{e,\mu}}{I_\tau}},\ \ \ I_i=
             \exp\left(\frac{1}{8\pi^2}\int^{\ln M_R}_{\ln M_Z}y^2_i 
             dt\right),
\end{eqnarray}
where $y_i$ is the Yukawa coupling. Thus, if the violations of $M^\nu_M(M_R)$ 
satisfy the condition Eq.~(18) at 
GUT scale and $\epsilon_i$ are small, it is clear that violation parameters 
$\Delta^\nu_i$ of $M^\nu_M(M_Z)$ also satisfy the condition Eq.~(18). 
Then the condition Eq.~(18) is stable with respect to radiative corrections.
     
\section{Discussions}
We tried to explain the quark and lepton mass hierarchies and small mixing 
of quarks and large mixing of neutrinos in the universal Yukawa coupling 
framework with small violations. We suppose $SU(5)$ as GUT at least because 
masses of the $d$ quark sector and the charged lepton sector are same order, 
and right-handed Majorana neutrino has to be introduced in order to explain 
the smallness of the neutrino mass compared to the charged leptons(see-saw 
mechanism).  For flavor symmetry, we did not assume any symmetry other than 
universality. In this work, we would search a rule for the violation parameters 
and find a symmetry for the flavor.
\par
For quark sector, we can get "power rule" for the violation parameters; 
$\delta^u_1=\lambda^8,\ \delta^u_-=\lambda^6,\ \delta^u_+=\lambda^4,\ 
\delta^d_1=\lambda^4,\ \delta^d_-=\lambda^3,\ \delta^d_+=\lambda^2,\ 
\lambda\approx0.32,\ \phi_+\equiv\phi\approx-4^\circ$. 
This very simple parameterization seems to give some suggestion to the flavor 
symmetry. For the lepton sector, it was shown that the condition 
${\rm Det}_2(\Delta^\nu)=(\Delta^\nu_2-\Delta^\nu_3-\Delta^\nu_5)^2
             -(\Delta^\nu_1-2\Delta^\nu_3)(\Delta^\nu_4-
             2\Delta^\nu_5)
             =0,\ 
{\rm Det}_3(\Delta^\nu)=2{\Delta^\nu_2}{\Delta^\nu_3}{\Delta^\nu_5}
             -{\Delta^\nu_1}{\Delta^\nu_5}^2-{\Delta^\nu_3}^2
             {\Delta^\nu_4}
             \ne0
$
for the violation parameters of the Majorana neutrino mass matrix must be 
satisfied for the large neutrino mixing. The stability of the condition to the 
radiative correction was shown. We can get solutions satisfying the charged
lepton and neutrino mass hierarchies and neutrino large mixing $V_{\rm MNS}$, 
but cannot get a rule for these violation parameters in this article. However, 
it is expected that an analysis following our scenario can find information 
for the family symmetry by analyzing the more precise data with respect to 
neutrino sector. 
         

\end{document}